Avalanches in directed complex networks of neurons connected by stochastic synapses

Short running title: Avalanches in directed stochastic networks of neurons

S.L. Ginzburg, M.A. Pustovoit

Petersburg Nuclear Physics Institute, Gatchina Leningrad region, 188300 Russia

Directed complex network of two-state model neurons linked by synapses which can be blocked or activated stochastically in time undergoes phase transition between the quiescent phase with zero activity and the active one with persistent activity. Avalanches of activity appear in the quiescent phase after occasional excitation of randomly chosen neuron. Size and duration distributions of long avalanches which we observe near phase transition point are exponential. Only a small fraction of all neurons are active at any moment of a long avalanche.

Recently, considerable attention was attracted to the phenomenon called "neuronal avalanches". This name was originally attributed to spontaneous activity of cultures of cortical neurons grown atop a regular array of electrodes ([1], see [2,3] for review). It appears that such activity occurs in the form of spatiotemporal patterns (network bursts) of electrode potentials separated by periods of quiescence. The distributions of sizes and durations of these avalanche-like events were measured and claimed to be of power-law type, with exponents typical to "critical branching process" [4]. These findings seem to support the hypothesis (pioneered by Turing [5] and stressed by Bak [6]) that brain tunes itself to the "edge of criticality" [7], where, for instance, it can respond to stimuli of wider dynamical range [8]. These studies are inspiring but demand a considerable amount of further work to understand functions and formation of avalanches, especially in relation to the structure of underlying network. In this respect, it is important that synapses that connect neurons in central nervous system are unreliable, that is, the postsynaptic response to an action potential generated by presynaptic neuron occurs stochastically, with probability much less than 1 [9]. Such behavior seems inexpedient from both computational and metabolic points of view, and its consequences for brain functioning are practically unknown.

The aim of present work is to construct a simple model of neural tissue (network) with stochastic synapses and study how the properties of avalanches reflect the underlying network structure and dynamics. Specifically, we construct a network with homogeneous (Gaussian) or heterogeneous (power-law or exponential) connectivity distribution, place two-state excitable elements (neurons) to nodes of the network, and allow the propagation of activity between them through directed edges (synapses) that can be blocked or activated stochastically in time. This representation is oversimplified both from the structural point of view (the connectivity distribution is a point of debate [7,10-13] even in small cultured networks) and from the dynamical one (we use discrete time and consider only excitatory interaction). Nevertheless, this toy model appears preferable for our purposes since it has only one control parameter, the probability of a synapse to transmit activity, and allows us to compare behavior of different networks in the region of phase transition. The dynamics of the model corresponds to the contact process widely used in various fields [14], so we suppose for it to be relevant, e.g., in epidemics studies.

We found that in all studied networks with occasional excitation of random node two types of avalanches occur. The short ones propagate locally, i.e., at the scale of average path length of the network. The distributions of their sizes and durations are similar to power law. The long avalanches can be orders of magnitude longer and demonstrate clear exponential distribution of sizes and durations. They frequently occur when the control parameter, the synaptic reliability, approaches threshold value above which they transform to persistent activity (endless avalanche). Thus we observe phase transition from the quiescent phase to the active one, and study the dynamics of network activity near the transition point.

We should stress that long avalanches occur not only in scale-free networks grown by linear preferential attachment of nodes but also in those grown without preference as well as in homogeneous random networks. This fact is especially important since the connectivity of real neural assemblies is largely unknown [10-13].

The samples of directed networks were prepared as follows. For growing network, we start from $n$ nodes connected randomly with directed edges. Next we add, one at a discrete time moment $t = 1, 2 \ldots$, new nodes with $m_{in}$ incoming and $m_{out}$ outgoing edges. The number $s$ of any node is the moment of its birth plus $n$. In every moment $t$ the $s$-th node has $k_{in}(s,t)$ incoming and $k_{out}(s,t)$ outgoing edges (the in-degree and the out-degree, accordingly). The probabilities of attachment of the new node's incoming or outgoing edge to the already existing $s$-th node are, for scale-free (SF) network, $p_{in}, p_{out} \sim k_{out}(s,t)$ (linear preferential attachment by out-degree), and, for exponential network, $p_{in} = p_{out} = 1/(n+t)$.

The total number of nodes in the obtained network is $M = n + t$. In the former case, the resulting SF network has an asymptotic power-law distribution of in- and out-degrees with the same exponent $\gamma = 2 + m_{out}/m_{in}$. In our simulations, we take $m_{in} = 14$, $m_{out} = 7$, $n = 35$, so $\gamma = 2.5$. In the latter case, the degree distributions are exponential. Finally, taking $n = M$, we obtain a homogeneous random network with Gaussian degree distribution.

The dynamic (depending on discrete time which we denote as $\tau$) quantities in our model are the states of all nodes and edges, which we consider as binary:

$$n_i(\tau) = \begin{cases} 0 \\ 1 \end{cases}, \quad J_{ij}(\tau) = \begin{cases} 0 \\ 1 \end{cases}, \tag{1}$$

where $J_{ij}$ represents the existing edge directed from node $j$ to node $i$. When $J_{ij} = 0$, the link is blocked, otherwise it is active. For nodes, $n_i = 0$ is the quiescent and $n_i = 1$ is the active (excited) state.

Any existing edge in any moment can be active with the probability $p$, i.e., $J_{ij}$ are random variables with distribution: $\varphi[J_{ij}(\tau)] = p\delta[J_{ij}(\tau)-1] + (1-p)\delta[J_{ij}(\tau)]$.

The dynamics of a node is described by the following algorithm:

1. If the node $i$ is quiescent at the moment $\tau$, it becomes active at the next moment $\tau + 1$ when, at the moment $\tau$, at least one active node acts on it through active directed edge;

2. If the node $i$ is active at the moment $\tau$, it becomes quiescent at the next moment $\tau + 1$.

Mathematically, the node dynamics is governed by the simple Markovian equation:

$$n_i(\tau+1) = [1 - n_i(\tau)]\theta\left[\sum_j J_{ij}(\tau) n_j(\tau)\right], \tag{2}$$

where $\theta(x)$ is the Heaviside theta function. Such dynamics of nodes corresponds to the SIS epidemic model or to the contact process that have been extensively studied in complex networks [14,15]. However, the stochastic activation or blockage of edges (annealed disorder) that reflects the behavior of synapses has not been considered in those models.

In our simulations we started from all nodes in a network being quiescent and activated a single randomly selected node. The resulting avalanche of activity propagated through the network for some time, after which, due to simultaneous blockage of all outgoing links of active nodes, the network became quiescent again, and the random perturbation was repeated. Such a non-steady stimulation was used in experiments with cortical cultures where neuronal avalanches were found [1].

The measure of the total activity of a network is the instantaneous fraction $\rho$ of active nodes. For $p \approx 1$ and sufficiently large $\langle k_{out} \rangle$ the activity lasts forever ($\rho > 0$). For very small $p$, on the contrary, the initial excitation of any node dies out quickly ($\rho = 0$). Thus a phase transition must exist at some intermediate value of $p$, and we study avalanches of finite size below (and near) this point. We define the duration $l$ of an avalanche as the time interval when $\rho > 0$. The commonly studied quantity, the avalanche size $P$ is the sum of values of $\rho$ for all time moments during an avalanche. Their cumulative distribution functions, $C(l)$ and $C(P)$, are shown in fig.1 for one sample of SF network with $M = 10^4$.

Thus we observe two distinct types of avalanches. The distributions of durations and sizes of the short ones ($l < 10$) are similar to power law but their small range does not allow to make a reliable analysis (see the data in the inset of fig.1(a) displayed by squares). We just note that similar distributions found in spontaneously active cortical cultures [1] served as the ground for revival of the hypothesis of critical state of brain networks ("brain at the edge of criticality") [7].

The avalanches of another type, the long ones, were, at our best knowledge, not studied before. They are exponentially distributed by both duration and size:

$$C(l) \sim \exp(-l/\lambda_1), \quad C(P) \sim \exp(-P/\lambda_2). \tag{3}$$

This is one of the main results of this paper. Exponential distribution of avalanche sizes was recently reported in vivo [16] and in simulations [17].

The characteristic duration $\lambda_1$ and size $\lambda_2$ increase exponentially (data not shown) when the probability of link activation $p$ approaches the phase transition point $p_c$. The latter can be found using the mean-field description similar to ref. [18], which results in the following formula for directed network:

$$p_c = \frac{\langle k_{out} \rangle}{\langle k_{out}^2 \rangle} \tag{4}$$

One can see that the threshold vanishes in an infinite SF network with $\gamma < 3$. For finite network size $M$, following the line of reasons presented in ref. [18] and using the "structural cutoff" $k_{c,out}^{(s)} = \sqrt{\langle k_{out} \rangle M}$ [19], we get for SF network:

$$p_c = \frac{3-\gamma}{\gamma-2} \cdot \frac{m_{out}^{2-\gamma} - (m_0 M)^{\frac{2-\gamma}{2}}}{(m_0 M)^{\frac{3-\gamma}{2}} - m_{out}^{3-\gamma}}, \quad m_0 \equiv m_{in} + m_{out}. \tag{5}$$

To verify eq.(5) numerically, we found the 10-samples average value of the threshold for networks of various sizes. Fig.2 displays the simulation results, which are in good agreement with eq.(5). The results for networks with homogeneous (Gaussian) and exponential out-degree distributions are presented as well, showing much larger $p_c$ for large $M$, in accordance with asymptotical values obtained from eq.(4).

The next important characteristic of an avalanche is its shape, i.e., the sequence of fractions of active elements at each time moment during an avalanche of duration $l$, $\rho_l(\tau), 1 \leq \tau \leq l$. The shape of one sufficiently long avalanche is shown in fig.3(a). One can see strong fluctuations of the number of active elements, which lead at last to the end of the avalanche for $p < p_c$. The meaningful picture can be obtained by averaging of shapes of all avalanches of the same duration. The results are shown in fig.3(b,c). We see now the main difference between short and long avalanches: the mean activity during the long ones reaches the universal plateau value $\rho_0$ and lasts there for a sufficiently long time. During this quasistationary state, only a small part of the whole network is active at a given moment. The fraction of nodes that are active is small even in the active phase not too far from $p_c$ (see fig.4). Such a low persistent activity is known as the endemic state in studies of epidemic spreading; it was occasionally observed in simulations of neural network models [20].

We study this quantity in detail since it reflects the dynamics of the order parameter in the quiescent-to-active phase transition. Indeed, in the active phase the order parameter is:

$$\rho = \langle \rho_\infty(\tau) \rangle, \qquad (6)$$

where the infinity sign means infinitely long avalanche, and the angle brackets denote averaging over the ensemble of networks. The order parameter is proportional to $-\varepsilon$, $\varepsilon = (p_c - p)/p_c$, in the active phase [18] and becomes zero in the quiescent one due to finite duration of all avalanches. However, due to finite network size $M$ there are fluctuations of order parameter which should be proportional to $M^{-1/2}$. These fluctuations should occur in both active and quiescent phases. In the latter one we have:

$$\langle \rho_l(\tau) \rangle = \rho_0(\varepsilon, M). \qquad (7)$$

The above considerations can be summarized as follows:

$$\rho_0(\varepsilon, M) = \begin{cases} M^{-1/2}(S - \varepsilon B_1(M)), & p > p_c, \\ M^{-1/2} S f(\varepsilon), & p < p_c. \end{cases} \qquad (8)$$

where $S$ is a constant, $B_1(M)$ is the (increasing) function of $M$, and $f(\varepsilon)$ should decay from unity at $\varepsilon = 0$ to zero at large $\varepsilon$ (i.e., far from the phase transition point).

The dependences of $\rho_0$ on $\varepsilon$ at both sides of phase transition point for networks of various sizes and both types are depicted in fig.4. The data collapse when $\rho_0$ is multiplied by $M^{1/2}$ clearly demonstrates the finite size scaling described by eq.(8). We see also that the linear dependence of avalanche height $\rho_0$ on $\varepsilon$ in the active phase changes to the exponential decay in the quiescent one, that is, $f(\varepsilon) = \exp(-\varepsilon B_2(M))$. The similar behavior of the order parameter was obtained in Ref. [21] for contact process on undirected weighted SF network. From fig.4 we can conclude that $B_2$ does not depend on $M$ for SF networks and $S \cong 2.3$ is the universal constant expected to depend on $m_{in}$ and $m_{out}$ only. Let us note also that long avalanches below $p_c$ are higher and their range of existence is wider for SF networks than for exponential ones. For homogeneous random networks we observed long avalanches only at $\varepsilon < 0.02$.

To conclude, we observed two types of avalanches of activity in networks of neurons connected by directed stochastic synapses. The short ones are local, i.e., they propagate at a distance of the order of network mean shortest path length (in our case it is $r_0 = \ln M / \ln(2m_0) = 2.5$ for $M = 10^4$ [15]). They are similar to "neuronal avalanches" observed previously. The long avalanches occur in the quiescent phase of various complex random networks (SF, non-SF small-world, and homogeneous) near the point of phase transition to active phase. Distributions of their durations and sizes decay exponentially. Nevertheless we obtain that they possess a large fraction of total observation time in a fairly wide range of the control parameter, especially for SF networks. The activity of a network during long avalanches fluctuates strongly around the mean value (avalanche height) most of the time. This mean activity increases with increase of the control parameter, but remains low even in the active phase. Thus, long avalanches in the quiescent phase and persistent activity in the active one resemble normal stochastic activity of brain structures rather than epilepsy.

The dynamics of our model correspond to the contact process well known in science, and we suppose that our results might be relevant, e.g., to epidemic spreading with unstable contacts between individuals.

The authors acknowledge financial support from Russian Foundation for Basic Research (Project 08-02-00314a) and the State Programs "Quantum Macrophysics", "Strongly Correlated Electrons in Metals, Superconductors, Semiconductors and Magnetic Materials", and "Neutron Studies of Matter".

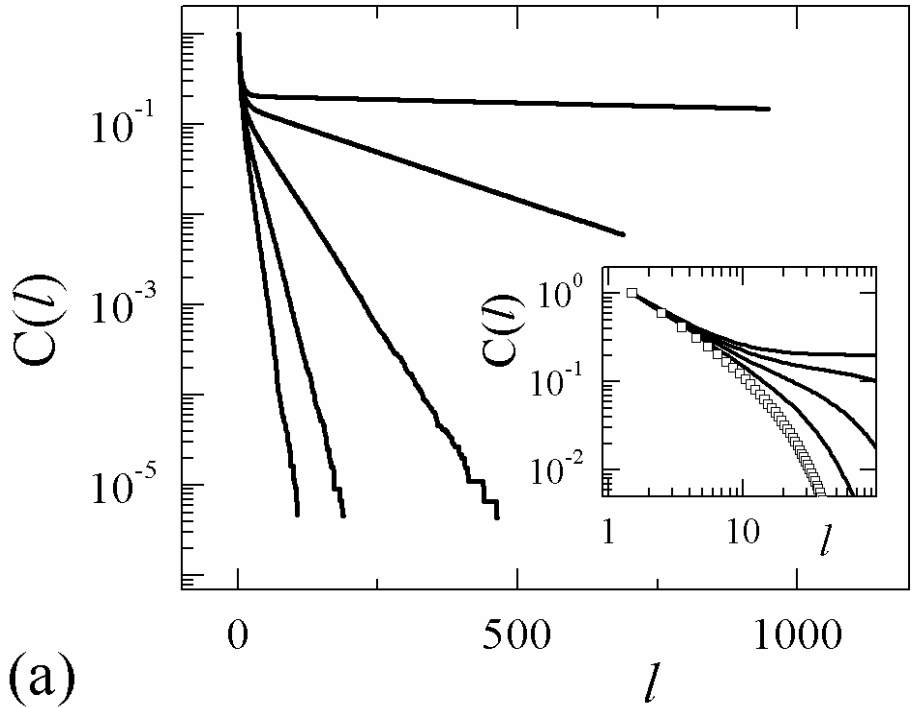

(a)

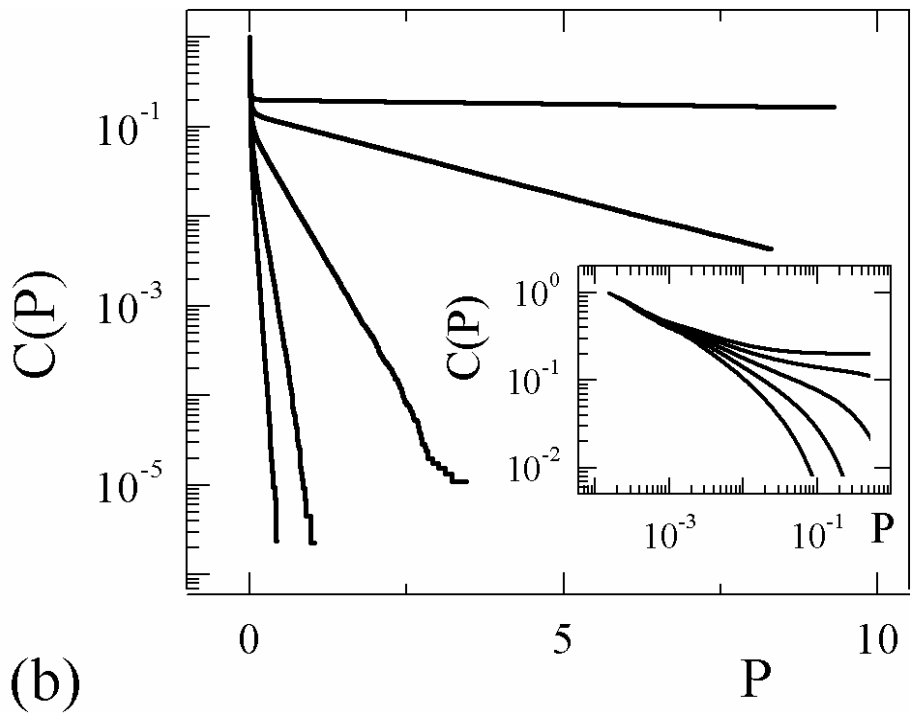

(b)

Fig.1. Cumulative distribution functions of avalanche durations (a) and sizes (b) for one sample of a network with $M = 10^4$. The probabilities of link activation are, from top to bottom: $p = 0.019, 0.018, ,0.017, 0.016, 0.015$. Inset shows the upper left portion of the graph in the log-log scale. One can see the exponential and powerlike distribution for large

and small avalanches, respectively. The phase transition point where endless avalanche appears is $p_c = 0.02$.

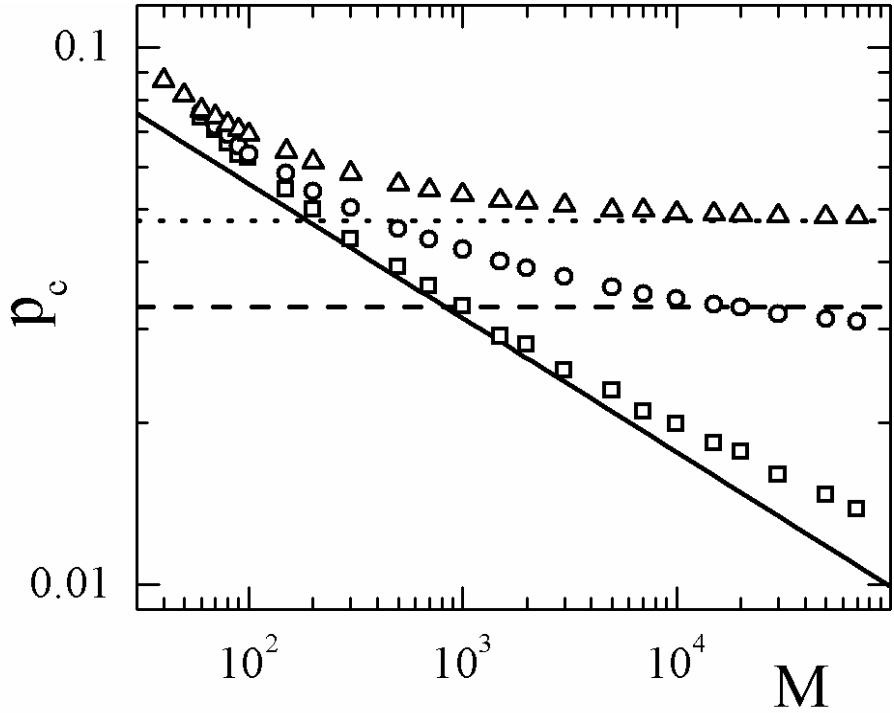

Fig.2. Dependence of the phase transition value of the probability of link activation on the network size for three types of degree distribution: scale-free with $\gamma = 2.5$ (squares), exponential (circles) and Gaussian (triangles). Datasets are obtained by averaging over 10 network samples. The solid curve corresponds to eq.(5), the dashed and dotted lines are, correspondingly, the asymptotical values for exponential, $p_c^{(Exp)} = m_0^2/(m_0^2 + m_{in}^2)$, and Gaussian, $p_c^{(G)} = 1/m_0$, network.

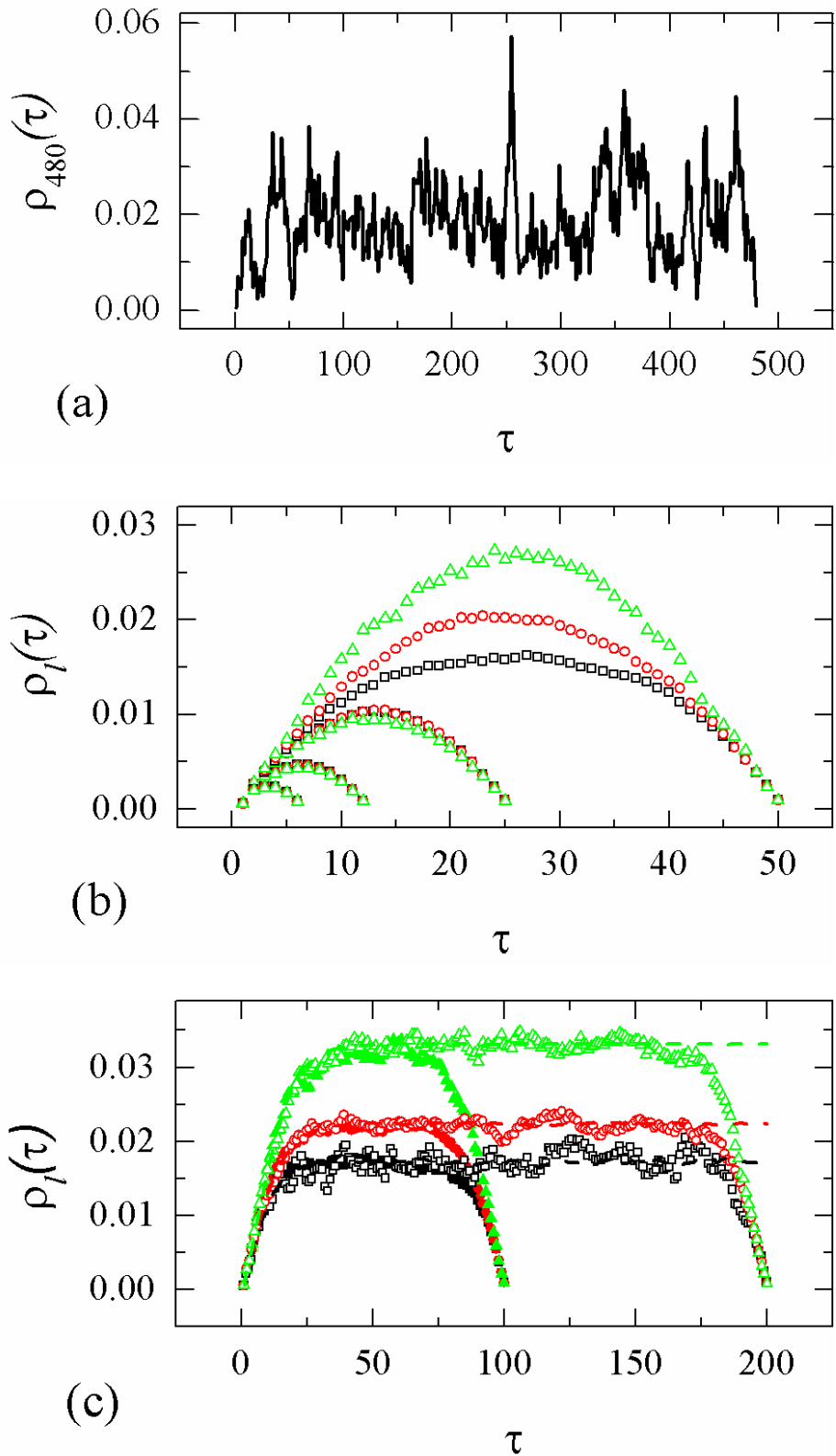

Fig.3. (a) Shape of the single long avalanche with duration $l = 480$ in the SF network with $M = 2500, p = 0.022$. (b),(c) Shapes of avalanches of various durations averaged by all avalanches of the same duration for $p = 0.022$ (squares), 0.023 (circles), and 0.0245 (up

triangles). Phase transition takes place at $p_c(2500) = 0.028$. The noisy character of averaged data is due to small number of avalanches of particular duration in the total collected $10^6$ avalanches. Smooth dashed curves in (c) are the averages over all avalanches with $l > 250$.

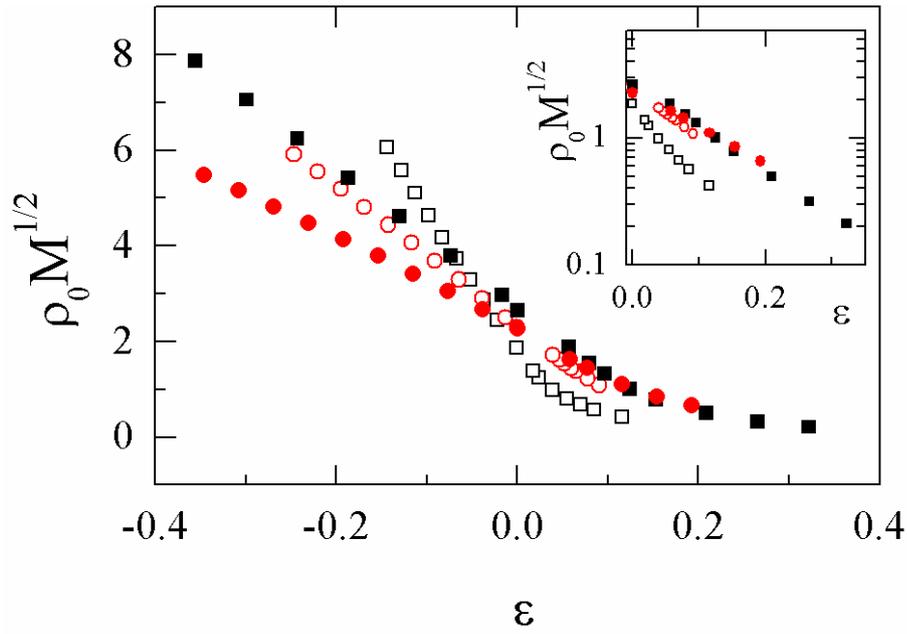

Fig.4. Avalanche height $\rho_0$ multiplied by $M^{1/2}$ in the vicinity of $p_c$ for networks of size $M = 2 \cdot 10^4$ (squares) and $2500$ (circles). The filled and open symbols correspond to networks with SF and exponential connectivity distribution, respectively. Inset: semilog plot of the region below $p_c$.